\documentstyle[aps,12pt]{revtex}
\begin{document}
\title{Relativistic Hartree-Bogoliubov description\\
of deformed light nuclei}
\author{G.A. Lalazissis$^{1,2}$, D. Vretenar$^{2,3}$, 
and P. Ring$^{2}$}
\bigskip
\address{
$^{1}$Department of Theoretical Physics, Aristotle University of Thessaloniki,
Thessaloniki GR-54006, Greece\\
$^{2}$ Physik-Department der Technischen Universit\"at M\"unchen,
D-85748 Garching, Germany\\
$^{3}$ Physics Department, Faculty of Science, University of
Zagreb, 10 000 Zagreb, Croatia\\}
\maketitle
\bigskip
\bigskip
\begin{abstract}
The Relativistic Hartree-Bogoliubov model is applied in 
the analysis of ground-state properties
of Be, B, C, N, F, Ne and Na isotopes. The model uses
the NL3 effective interaction in the mean-field Lagrangian,
and describes pairing correlations
by the pairing part of the finite range Gogny interaction D1S.
Neutron separation energies, quadrupole deformations, nuclear matter radii,
and differences in radii of proton and neutron distributions are
compared with recent experimental data.
%#####################################################
\end{abstract}
%#####################################################
\bigskip \bigskip

%#####################################################
\vspace{1 cm} {PACS:} {21.60.Jz, 21.10.Dr, 21.10.Gv, 
27.20.+n, 27.30.+t}\newline
\vspace{1 cm}\newline
\newpage
\baselineskip = 24pt

%=========================================================================
%  Section 1
%
\section{Introduction and Outline of the Model}
%=========================================================================
During the last decade a large quantity of data on light nuclei
with $4\leq Z\leq 12$ has become available. In particular, 
measurements of interaction cross sections by using 
radioactive nuclear beams at intermediate and relativistic 
energies, have provided important data on nuclear
radii~\cite{Tan.88,Oza.94,Suz.95,Oza.96,Suz.98,Suz.99,Oza.01}.
The nuclear radius is a fundamental quantity which, in principle,
provides information on the effective nuclear potential, 
shell effects and ground-state deformation. For exotic 
nuclei with extreme values of neutron to proton ratio, 
particularly important is the isospin dependence of nuclear
radii which can signal the onset of new phenomena like, 
for example, the formation of skin and halo structures.
Data on ground-state deformations are also very important
for the study of shell effects
in exotic nuclei. In particular, they reflect major 
modifications in the shell structures, the disappearance
of standard and the occurrence of new magic numbers. Different
ground-state deformations of proton and neutron density 
distributions are expected in some nuclei with extreme 
isospin projection quantum number $T_z$. Another source 
of information on the effective nuclear potential 
in exotic systems at the limits of stability
are the single-nucleon 
separation energies. The neutron drip-line has been 
reached for nuclei up to $Z=9$~\cite{Sak.99}. On the 
proton-rich side the drip-line has been experimentally 
fully mapped up to Z=21, and
possibly for odd-Z nuclei up to In \cite{WD.97}.
In very neutron-rich nuclei
the weak binding of the outermost neutrons causes 
the formation of the neutron skin on the surface of a nucleus,
and the formation of one- and two-neutron halo structures.
The established two-neutron halo nuclei are $^6$He, $^{11}$Li, 
and $^{14}$Be, and the one-neutron halo nuclei are 
$^{11}$Be and $^{19}$C. Recent data~\cite{Oza.01} present 
evidence for a one-neutron halo in $^{22}$N, $^{23}$O and
$^{24}$F. The formation of the neutron skin is well 
established in the neutron-rich Na isotopes~\cite{Suz.95,Suz.98},
and the related phenomenon of low-lying pygmy
isovector dipole resonances has recently been observed 
in O isotopes~\cite{Leist.01}. On the proton-rich side
evidence has been reported for the existence of a proton skin
in $^{20}$Mg~\cite{Chu.96}, and a beautiful example of 
exotic decay modes is provided by the two-proton emitter
$^{18}$Ne~\cite{Cam.01}.

In the present work the Relativistic Hartree-Bogoliubov (RHB)
model is applied in the analysis of ground-state properties
of Be, B, C, N, F, Ne and Na isotopes.
Based on the relativistic mean-field theory and on the 
Hartree-Fock-Bogoliubov framework, the RHB model provides a 
unified description of mean-field and pairing correlations.
It has been successfully applied in the description of 
nuclear structure phenomena in exotic nuclei far from the valley
of $\beta$- stability and of the physics of the drip lines.
On the neutron-rich side RHB studies include:
the halo phenomenon in light nuclei~\cite{PVL.97},
properties of light nuclei near the neutron-drip~\cite{LVP.98},
the reduction of the spin-orbit potential in nuclei with extreme
isospin values~\cite{LVR.97},
the deformation and shape coexistence phenomena that 
result from the suppression
of the spherical N=28 shell gap in neutron-rich nuclei~\cite{Lal.99},
properties of neutron-rich Ni and Sn isotopes~\cite{LVR.98}.
In proton-rich nuclei the RHB model has been used
to map the drip line 
from $Z=31$ to $Z=73$, and the phenomenon of 
ground-state proton radioactivity
has been studied~\cite{VLR.99,LVR.99,LVR.01b}.
In a very recent study of the isovector channel
of the RHB model~\cite{LVR.01}, a very good agreement with experimental
data has been obtained for
ground-state properties of nuclei that 
belong to the $A = 20$ isobaric sequence.

In the framework of the relativistic mean field (RMF)
approximation~\cite{Rin.96}
nucleons are described as point particles that
move independently in the mean fields
which originate from the nucleon-nucleon interaction.
The theory is fully Lorentz invariant.
Conditions of causality and Lorentz invariance impose that the
interaction is mediated by the
exchange of point-like effective mesons, which couple to the nucleons
at local vertices. The single-nucleon dynamics is described by the
Dirac equation
\begin{equation}
\label{statDirac}
\left\{-i\mbox{\boldmath $\alpha$}
\cdot\mbox{\boldmath $\nabla$}
+\beta(m+g_\sigma \sigma)
+g_\omega \omega^0+g_\rho\tau_3\rho^0_3
+e\frac{(1-\tau_3)}{2} A^0\right\}\psi_i=
\varepsilon_i\psi_i.
\end{equation}
$\sigma$, $\omega$, and
$\rho$ are the meson fields, and $A$ denotes the electromagnetic potential.
$g_\sigma$ $g_\omega$, and $g_\rho$ are the corresponding coupling
constants for the mesons to the nucleon.
The lowest order of the quantum field theory is the {\it
mean-field} approximation: the meson field operators are
replaced by their expectation values. The sources
of the meson fields are defined by the nucleon densities
and currents.  The ground state of a nucleus is described
by the stationary self-consistent solution of the coupled
system of Dirac and Klein-Gordon equations.

In addition to the self-consistent mean-field
potential, pairing correlations have to be included in order to
describe ground-state properties of open-shell nuclei.
In the framework of the
relativistic Hartree-Bogoliubov model,
the ground state of a nucleus $\vert \Phi >$ is represented
by the product of independent single-quasiparticle states.
These states are eigenvectors of the
generalized single-nucleon Hamiltonian which
contains two average potentials: the self-consistent mean-field
$\hat\Gamma$ which encloses all the long range particle-hole ({\it ph})
correlations, and a pairing field $\hat\Delta$ which sums
up the particle-particle ({\it pp}) correlations.
In the Hartree approximation for
the self-consistent mean field, the relativistic
Hartree-Bogoliubov equations read
\begin{eqnarray}
\label{equ.2.2}
\left( \matrix{ \hat h_D -m- \lambda & \hat\Delta \cr
		-\hat\Delta^* & -\hat h_D + m +\lambda} \right)
\left( \matrix{ U_k({\bf r}) \cr V_k({\bf r}) } \right) =
E_k\left( \matrix{ U_k({\bf r}) \cr V_k({\bf r}) } \right).
\end{eqnarray}
where $\hat h_D$ is the single-nucleon Dirac
Hamiltonian (\ref{statDirac}), and $m$ is the nucleon mass.
The chemical potential $\lambda$  has to be determined by
the particle number subsidiary condition in order that the
expectation value of the particle number operator
in the ground state equals the number of nucleons.
$\hat\Delta $ is the pairing field. The column
vectors denote the quasi-particle spinors and $E_k$
are the quasi-particle energies.

The RHB equations are solved self-consistently, with
potentials determined in the mean-field approximation from
solutions of Klein-Gordon equations for the meson fields.
The Dirac-Hartree-Bogoliubov equations and the equations for the
meson fields are solved by expanding the nucleon spinors
$U_k({\bf r})$ and $V_k({\bf r})$,
and the meson fields in terms of the eigenfunctions of a
deformed axially symmetric oscillator potential.
The calculations for the present analysis have been performed
by an expansion in 14 oscillator shells for the fermion fields,
and 20 shells for the boson fields.
A simple blocking procedure is used in the calculation of
odd-proton and/or odd-neutron systems. The blocking calculations
are performed without breaking the time-reversal symmetry.
A detailed description of the Relativistic Hartree-Bogoliubov
model for deformed nuclei can be found in Ref. \cite{LVR.99}.
%=========================================================================
%  Section 2
%
\section{Ground-state properties of deformed light nuclei}
%=========================================================================
In parallel with the experimental work of the last decade,
many theoretical analyses have been 
performed of the structure of nuclei in the mass region 
$10\leq A\leq 30$. Both microscopic mean-field and shell-model
approaches, as well as various microscopic cluster models, 
have been used to study properties of ground and excited 
states of isotopic and isobaric sequences, and to describe specific
structure phenomena in exotic nuclei. It has been 
shown that theoretical models reproduce the global trends 
of nuclear sizes and binding energies. However, 
special assumptions have to be made, or even new models have 
to be designed, in order to describe more exotic phenomena like,
for example, the location of the neutron drip line in Oxygen, or
the ground-state deformation of $^{32}$Mg.

In Ref.~\cite{LVP.98} we reported spherical RHB calculations 
of neutron-rich isotopes of N, O, F, Ne, Na and Mg. 
By using several standard RMF effective interactions, 
we analyzed the location
of the neutron drip-line, the reduction of the spin-orbit interaction,
$rms$ radii, changes in surface properties, and the formation of
neutron skins and of neutron halos. It was shown that, 
even without taking into account the deformation of the mean field,
the RHB model correctly describes the global trends of the
observed ground-state properties. The exception is, of course,
the location of the neutron drip line in Oxygen, which none
of the RMF effective interactions reproduces.

In the study of Ref.~\cite{LVR.01} we performed deformed
RHB calculations of ground-state properties of nuclei that
belong to the $A = 20$ isobaric sequence. 
The NL3 effective interaction~\cite{LKR.97}
was used for the mean-field Lagrangian, and pairing correlations
were described by the pairing part of the finite range Gogny 
interaction D1S~\cite{BGG.84}. 
This particular combination of effective forces 
in the $ph$ and $pp$ channels has been used in most of our 
recent applications of the RHB theory. RHB results for
binding energies, 
neutron and proton ground-state density distributions,
quadrupole deformations, nuclear matter radii, and proton radii
were compared with available experimental data.
The very good agreement with the observed ground-state properties 
as function of the isospin projection $T_z$, led to the conclusion 
that the isovector channel
of the NL3 interaction is correctly parameterized
and that this effective force can be used to describe not only
medium-heavy and heavy nuclei~\cite{LVR.98,VLR.99,LVR.99,LVR.01b}, but also 
properties of relatively light nuclei far from $\beta$-stability.

In the present work we apply the RHB model, with the NL3+D1S
effective interaction, in the analysis of ground state properties
of Be, B, C, N, F, Ne and Na isotopic sequences. We perform 
deformed RHB calculations and compare radii, separation 
energies and quadrupole deformations with available 
experimental data and with the predictions of the finite range
droplet model (FRDM)~\cite{MN.95}. Since the RHB equations are solved
in the configuration space of harmonic oscillator basis states,
for nuclei at the drip lines we do not expect an accurate
description of properties that crucially depend on the
spatial extension of the wave functions of the outermost
nucleons, especially on the neutron-rich side. Thus, we
do not attempt to describe radii of halo nuclei. We also
do not repeat the spherical calculations of O isotopes, 
which can be found in Ref.~\cite{LVP.98}.

In Fig.~\ref{figA} we display the proton, neutron and 
matter radii, ground-state quadrupole deformations,
and one-neutron separation energies
of Beryllium isotopes, calculated
with the NL3 + Gogny D1S effective interaction.
The RHB values are compared with the experimental
radii~\cite{Tan.88,Suz.99} and separation energies~\cite{AW.95}.
The calculated matter radii reproduce the trend of the experimental values,
except for the halo nucleus $^{11}$Be~\cite{Fuk.91,Kell.95}.
For the proton radii, on the other hand, the theoretical
values are considerably lower than the experimental ones~\cite{Tan.88},
especially for $^{14}$Be, though it should be noted that
the experimental proton radius has a very large uncertainty. 
The calculated neutron skin in $^{14}$Be is very large: 
$r_n - r_p = 0.71$ fm, and also the deformations of the 
proton and neutron densities in this nucleus are very 
different. $^{14}$Be has a large prolate deformation
$\beta_2 = 0.36$ and the ratio of neutron to proton quadrupole
moments is $Q_n / Q_p = 3.6$. The one-neutron separation energies are also
in excellent agreement with the empirical values~\cite{AW.95}. 
In particular, $^{13}$Be is predicted to be unbound by 180 keV.

In Fig.~\ref{figB} the same comparison is shown for neutron-rich
Boron isotopes. The calculated matter radii are in excellent
agreement with the recent experimental data~\cite{Suz.99}
for $^{17}$B and $^{19}$B, while they are larger than the 
older experimental values~\cite{Tan.88} for $^{14}$B and $^{15}$B.
Unlike in the case of Be, the calculated proton radii for 
these two nuclei agree well with the empirical values, but the 
theoretical neutron radii are much larger. The RHB calculation
also predicts $^{14}$B and $^{15}$B to be spherical in the 
ground state, while the heavier Boron isotopes are strongly
prolate deformed. The separation energies agree with the 
empirical values~\cite{AW.95}, and we note that both 
$^{16}$B and $^{18}$B are predicted to be neutron unbound. 

The calculated quantities which characterize the ground-states of
neutron-rich Carbon isotopes are displayed in Fig.~\ref{figC}.
The proton radius of $^{14}$C is compared with the experimental value 
from Ref.~\cite{Vri.87}, and the matter radii are shown 
in comparison with very recent experimental data~\cite{Oza.01}.
The trend of the experimental matter radii is clearly 
reproduced by the RHB calculation. Of course, for the one-neutron
halo nucleus $^{19}$C~\cite{Baz.95} the calculation in the harmonic 
oscillator configuration space cannot reproduce the anomalous matter
radius. The RHB model predictions for the one-neutron separation 
energies are in agreement with the empirical values ~\cite{AW.95},
though the theoretical values are slightly larger for the even-N 
isotopes. In particular, from an analysis of the angular distribution
of the $^{18}$C~$+ n$ center of mass~\cite{Nak.99}, the neutron 
separation energy is determined to be $530\pm 130$ keV, and 
the RHB calculated value is 510 keV.
$^{14}$C and $^{22}$C are spherical, and all other neutron-rich 
Carbon isotopes, except $^{17}$C, have oblate ground states. 
The prolate minima, however, are found at relatively low 
excitation energies, and
in Fig.~\ref{figC} we also plot the energy differences between
prolate and oblate minima. 

The RHB results for Nitrogen isotopes are shown in Fig.~\ref{figD}.
The proton radii are compared with the experimental values
from Ref.~\cite{Vri.87}. The calculated matter radii reproduce
the global trend of the experimental data~\cite{Oza.01}, but not
the sudden increase of the radii at N=15. In fact, in the recent
measurement of the interaction cross sections for $^{14-23}$N,
$^{16-24}$O, and  $^{18-26}$F on Carbon targets at relativistic
energies~\cite{Oza.01}, a sharp increase of matter radii at N=15 was
observed in all three isotopic chains (see also Fig.~\ref{figF}).
The deduced matter radii for $^{22}$N, $^{23}$O, and 
$^{24}$F are much larger than those of their respective neighbors
with one neutron less, 
and the deduced nucleon density distributions show a long 
neutron tail for these nuclei, comparable to that in $^{11}$Be.
It was therefore concluded that these three nuclei exhibit a 
one-neutron halo structure. Since they are spherical, the halo
structure should result from the odd neutron occupying the 
$2s_{1/2}$ orbital. The absence of the centrifugal barrier
for an $s$-orbital facilitates the formation of the long 
tail of the wave function, i.e. the halo structure. However,
in Ref.~\cite{Oza.01} it was also noted that the one-neutron
separation energies for $^{22}$N, $^{23}$O, and $^{24}$F 
are larger than 1 MeV ($1.22\pm 0.22$ MeV, $2.74\pm 0.12$ MeV 
and $3.86\pm 0.11$ MeV, respectively), and much larger than 
in $^{11}$Be and $^{19}$C. In a very recent analysis~\cite{Kan.01}
it has been pointed out that the conventional fixed core-plus-neutron
model for halo nuclei is unable to explain the observed enhanced 
cross section for these nuclei by any selection of neutron 
orbitals, and therefore a possibility of an enlarged core structure
has been suggested. The mechanism of core enlargement in these three
nuclei, however, has not been explained. We first note that the 
present RHB calculation reproduces the empirical one-neutron separation
energies ~\cite{AW.95}. In particular, for $^{22}$N we even obtain 
a slightly lower one-neutron separation energy, and the theoretical
matter radius coincides with the one deduced from the experimental
interaction cross section. Also for $^{23}$N, the calculated and
empirical separation energies coincide, and the theoretical
matter radius is only slightly below the large experimental 
error bar. The main difference is in the matter radii
of the lighter isotopes (a similar situation also occurs for 
the Fluorine isotopes, see Fig.~\ref{figF}). The calculated 
radii are somewhat larger than the experimental values and 
therefore at N=15 do not display the sharp discontinuity which, 
in Ref.~\cite{Oza.01}, is taken as evidence for the formation 
of the neutron halo. In the present calculation the gradual 
increase of matter radii reflects the formation of the neutron skin.
This is shown in the upper right panel of Fig.~\ref{figD}, where
the values of $r_n - r_p$ are plotted as function of the 
neutron number, and in Fig.~\ref{figE} we display 
the self-consistent RHB neutron and proton ground-state density
distributions of even-N Nitrogen isotopes. It should be pointed
out, however, that the formation of the halo structure 
can only be observed if calculations were performed in coordinate
space. Moreover, particle number projection might be necessary 
in order to reproduce the sharp increase of matter radii.
Finally, we note that the RHB NL3+D1S calculation predicts
the heaviest particle stable Nitrogen
isotope to be $^{23}$N, in excellent agreement with recent data
on the neutron drip-line~\cite{Sak.99}. 
  
Very similar results are obtained for the Fluorine isotopes.
In Fig.~\ref{figF} we compare the RHB theoretical
proton, neutron and matter radii, and one-neutron separation energies
with the experimental radii~\cite{Oza.94,Oza.01,Vri.87} and
separation energies~\cite{AW.95}. The calculated matter radii
do not reproduce the discontinuity at N=15, though for the 
heaviest isotopes they are found within the experimental
error bars. We note, however, that in the case of Fluorine
isotopes the calculated one-neutron separation energies do 
not reproduce the staggering between even-N and odd-N values
below $N\leq 15$. It is interesting to note that, like in the
case of Nitrogen, the RHB model with the NL3+D1S effective
interaction correctly predicts the location of the 
drip-line~\cite{Sak.99}: the last bound isotope of Fluorine
is $^{31}$F. Therefore, in agreement with experimental data,
we obtain that the neutron drip-line is at N=16 for Z=7, 
and at N=22 for Z=9. On the other hand, none of the standard
RMF effective interactions reproduces the location of 
the drip-line for Oxygen. It has been argued that the sudden
change in stability from Oxygen to Fluorine may indicate
the onset of deformation for the neutron-rich Fluorine
isotopes~\cite{Sak.99}. In the present calculation, however,
all Fluorine isotopes up to $^{31}$F turn out to be essentially
spherical.

In Ref.~\cite{PVL.97} we performed spherical RHB calculations
of the Ne isotopic chain. In particular, we studied the 
formation of neutron halo structures in drip-line Ne nuclei
($N > 20$). It was shown that the properties of the 1f-2p
neutron orbitals near the Fermi level, and the 
neutron pairing interaction play a crucial role in the 
possible formation of the multi-neutron halo. In the present
analysis we have performed calculations in the deformed 
harmonic oscillator basis. In Fig.~\ref{figG} we display the 
proton, neutron and matter radii, the ground-state quadrupole
deformations and the one-neutron separation energies 
of Ne isotopes.  
The calculated $\beta_2$ values are shown in comparison with the 
predictions of the finite-range droplet model~\cite{MN.95},
and the separation energies are compared with experimental
data~\cite{AW.95}. The FRDM and the present RHB calculation
predict a similar mass dependence of the ground-state 
quadrupole deformation. We note two spherical regions 
around A=16 and A=28. Both models reproduce 
the large prolate deformations around A=20, and predict 
prolate shapes in the region of possible halo structures
$A\geq 30$. Pronounced differences in the predicted
$\beta_2$ values are found for A=19 and A=24,25.
The later probably indicates a region of shape coexistence.
For A=28 the FRDM predicts
an oblate $\beta \approx -0.2$ deformation, while a spherical
shape is calculated in the RHB model. 
The calculated separation energies reproduce
the odd-even staggering and agree quite well with the 
experimental values.

The ground-state properties of the Na isotopic sequence are
illustrated in Fig.~\ref{figH}. The one-neutron separation energies
are shown in comparison with experimental data~\cite{AW.95}.
The calculated values reproduce the empirical staggering 
between even- and odd-A isotopes, although for $A > 24$ 
the theoretical separation energies are systematically 
somewhat larger for the even-N isotopes. The calculated
radii are compared with the experimental data: 
matter radii~\cite{Suz.98},
neutron radii~\cite{Suz.95}, and charge isotope 
shifts~\protect\cite{Ott.89}. An excellent agreement between 
theory and experiment is obtained. For the matter and neutron
radii the only significant difference is at A=22, but this 
dip in the experimental sequence has recently been attributed 
to an admixture of the isomeric state in the beam~\cite{Suz.98}.
Except for the lightest isotope shown, i.e. $^{20}$Na,
the calculated charge isotope shifts reproduce the 
empirical A-dependence. A significant difference between
the theoretical and experimental values is observed only
for $A \geq 29$. The calculated ground-state quadrupole
deformations of the Na isotopes are compared with the 
predictions of the finite-range droplet
model~\cite{MN.95}. We note that, while the FRDM predicts
all Na isotopes with $A\leq 28$ to be strongly prolate deformed,
the result of the RHB calculation is the staggering between
prolate and oblate shapes, indicating the onset of shape
coexistence. In particular, $^{26,27}$Na are predicted to be
oblate, while prolate ground-state deformations are 
calculated for $^{28,29}$Na. Very 
recent experimental data on quadrupole moments of
$^{26-29}$Na~\cite{Keim.00} confirm this prediction.

Our results for matter radii are summarized in Fig.~\ref{figI},
where we plot the calculated values for 
the $A = 14, 15, 16, 17, 18, 19$ isobaric chains as
functions of the isospin projection $T_z$. An excellent
overall agreement is found between the experimental data
and the matter radii calculated with the NL3+D1S RHB
effective interaction. 

Finally, in Fig.~\ref{figJ} the matter radii of mirror nuclei
are compared as function of the isospin projection $T_z$.
It is interesting to note that for $\Delta T_z = 1$, 
the nuclei with $T_z > 0$ have larger radii than their
mirror partners with $T_z < 0$. Due to the strong effect
of the Coulomb interaction, on the other hand, for
$\Delta T_z = 2$ and 3 the proton-rich nuclei have 
almost 0.1 fm larger nuclei than their $T_z > 0$ mirror 
partners.
\bigskip

%=========================================================================
%  Section 3
%
\section{Summary}
%=========================================================================
This work presents an analysis of ground-state properties
of Be, B, C, N, F, Ne and Na isotopes in the framework
of the Relativistic Hartree-Bogoliubov (RHB) model. In the 
last couple of years this model has been very successfully 
applied in the description of 
nuclear structure phenomena in medium-heavy and heavy exotic
nuclei far from the valley
of $\beta$- stability and of the physics of the drip lines.
The present analysis covers a region which is probably
at the low-mass limit of applicability of the mean-field 
framework. This work is also a continuation of our 
previous applications of the RHB model of Ref.~\cite{LVP.98}
(spherical RHB calculations of neutron-rich isotopes of N, O,
F, Ne, Na and Mg), and of Ref.~\cite{LVR.01} 
(deformed RHB calculations of ground-state properties of
the $A = 20$ isobaric sequence).

The present calculation has been performed in the 
configuration space of a deformed harmonic oscillator 
basis states. The NL3 effective interaction has been 
used for the mean-field Lagrangian, and pairing correlations
have been described by the pairing part of the finite range Gogny 
interaction D1S. The calculated
neutron separation energies, quadrupole deformations, nuclear matter radii,
and differences in radii of proton and neutron distributions have been
compared with very recent experimental data. Here we summarize the main
results:
\begin{itemize}
\item For the neutron drip-line nucleus $^{14}$Be the RHB calculation 
predicts a large prolate deformation $\beta_2 = 0.36$ and the ratio
of neutron to proton quadrupole moments $Q_n / Q_p = 3.6$.
\item For the neutron-rich Boron isotopes, the calculated matter
radii reproduce 
the recent experimental data~\cite{Suz.99}
for $^{17}$B and $^{19}$B. This is an important result, since
$^{19}$B ($T_z = 9/2$) has one of the largest N/Z values known
at present in the low-mass region of the nuclear chart.
\item Although the present calculation, performed in the deformed harmonic
oscillator configuration space, cannot reproduce the anomalous matter 
radius of the one-neutron halo nucleus $^{19}$C, the neutron 
separation energy 510 keV is in excellent agreement with the 
experimental value $530\pm 130$ keV obtained from an analysis of the
angular distribution of the $^{18}$C~$+ n$ center of mass~\cite{Nak.99}.
A large oblate deformation $\beta_2 \approx -0.4$ is predicted for $^{19}$C.
\item The RHB model with the NL3+D1S effective interaction predicts
the location of the neutron drip-line in Nitrogen and Fluorine in agreement
with recent experimental findings~\cite{Sak.99}:
the heaviest particle stable isotopes 
are $^{23}$N and $^{31}$F. The calculation, however,
does not reproduce the location of the neutron drip line in Oxygen.
\item The calculated matter radii of the neutron-rich Nitrogen
and Fluorine isotopes are in agreement with very recent experimental 
data~\cite{Oza.01}. However, the calculation does not reproduce the 
sudden increase of the radii at N=15, which was taken as evidence for 
the formation of the neutron halo.
\item For the neutron-rich Neon isotopes the RHB model predicts
pronounced prolate deformations in the region of possible halo structures
($A > 30$).
\item For the Na isotopic sequence the calculated radii are in excellent
agreement with experimental data on matter radii~\cite{Suz.98},
neutron radii~\cite{Suz.95}, and charge isotope
shifts~\protect\cite{Ott.89}. The calculated ground-state quadrupole
deformations confirm the recent experimental data on quadrupole moments of
$^{26-29}$Na~\cite{Keim.00}.
\end{itemize}

\bigskip
\bigskip

\leftline{\bf ACKNOWLEDGMENTS}

This work has been supported in part by the
Bundesministerium f\"ur Bildung und Forschung under
project 06 TM 979, and by the Gesellschaft f\" ur
Schwerionenforschung (GSI) Darmstadt.
D.V. would like to acknowledge the support from the Alexander von
Humboldt - Stiftung.

\newpage

\begin{figure}
\caption{
Proton, neutron and matter radii, ground-state quadrupole 
deformations and one-neutron separation energies 
of Beryllium isotopes, calculated
with the NL3 + Gogny D1S effective interaction.
The theoretical values are compared with the experimental
radii~\protect\cite{Tan.88,Suz.99} and
separation energies~\protect\cite{AW.95}.}
\label{figA}
\end{figure}

\begin{figure}
\caption{Same as in Fig.~\protect\ref{figA}, but for
Boron isotopes.}
\label{figB}
\end{figure}

\begin{figure}
\caption{
Proton, neutron and matter radii, ground-state quadrupole
deformations, one-neutron separation energies and
energy differences between prolate and oblate 
minima for Carbon isotopes.  
The theoretical values are compared with the experimental
radii~\protect\cite{Vri.87,Oza.01} and
separation energies~\protect\cite{AW.95}.}
\label{figC}
\end{figure}

\begin{figure}
\caption{
The RHB theoretical
proton, neutron and matter radii, and one-neutron separation energies
of Nitrogen isotopes, compared
with the experimental radii~\protect\cite{Vri.87,Oza.94,Oza.01} and
separation energies~\protect\cite{AW.95}.}
\label{figD}
\end{figure}

\begin{figure}
\caption{
Self-consistent RHB neutron and proton density 
profiles of Nitrogen isotopes.} 
\label{figE}
\end{figure}
 
\begin{figure}
\caption{
The RHB theoretical
proton, neutron and matter radii, and one-neutron separation energies
of Fluorine isotopes, compared
with the experimental radii~\protect\cite{Vri.87,Oza.94,Oza.01} and
separation energies~\protect\cite{AW.95}.}
\label{figF}
\end{figure}

\begin{figure}
\caption{
Proton, neutron and matter radii, ground-state quadrupole
deformations and one-neutron separation energies 
of Neon isotopes.  
The calculated $\beta_2$ values are displayed in comparison with the 
predictions of the finite-range droplet model~\protect\cite{MN.95},
and the separation energies are compared with experimental
data~\protect\cite{AW.95}.}
\label{figG}
\end{figure}

\begin{figure}
\caption{
One-neutron separation energies, radii, charge isotope shifts
and ground-state quadrupole deformations of Sodium isotopes.
The RHB calculated values are compared with the experimental 
data: one-neutron separation energies~\protect\cite{AW.95},
matter radii~\protect\cite{Suz.98},
neutron radii~\protect\cite{Suz.95}, charge isotope 
shifts~\protect\cite{Ott.89}, and with the $\beta_2$ values
calculated in the finite-range droplet
model~\protect\cite{MN.95}.}
\label{figH}
\end{figure}

\begin{figure}
\caption{
The nuclear matter radii of the $A = 14, 15, 16, 17, 18, 19$
isobaric chains as
functions of the isospin projection $T_z$. Results of fully
self-consistent RHB calculations are compared with experimental
data (for description see the text).}
\label{figI}
\end{figure}

\begin{figure}
\caption{
Comparison of calculated matter radii for mirror nuclei
as function of the isospin projection $T_z$.}
\label{figJ}
\end{figure}

%%%%%%%%%%%%%%%%%%%%%%%%%%%%%%%%%%%%%%%%%%%%%%%%%%%%%%%%%%%%%

\newpage
\bigskip
\begin{thebibliography}{999}


\bibitem{Tan.88} I. Tanihata {\it et al.}, 
	Phys. Lett. {\bf B206}, 592 (1988).
\bibitem{Oza.94} A. Ozawa {\it et al.},
	Phys. Lett. {\bf B334}, 18 (1994).
\bibitem{Suz.95} T. Suzuki {\it et al.}, Phys. Rev. Lett. {\bf 75},
	3241 (1995).
\bibitem{Oza.96} A. Ozawa {\it et al.},
	 Nucl. Phys. {\bf A608}, 63 (1996).
\bibitem{Suz.98}  T. Suzuki {\it et al.},
	 Nucl. Phys. {\bf A630}, 661 (1998).
\bibitem{Suz.99} T. Suzuki {\it et al.}, 
	Nucl. Phys. {\bf A658}, 313 (1999).
\bibitem{Oza.01} A. Ozawa {\it et al.},
	Nucl. Phys. {\bf A691}, 599 (2001).
\bibitem{Sak.99} H. Sakurai {\it et al.},
	 Phys. Lett. {\bf B448}, 180 (1999).
\bibitem{WD.97}  P.J. Woods and C.N. Davids, Annu. Rev. Nucl. Part. Sci.
	{\bf 47}, 541 (1997).
\bibitem{Leist.01} A. Leistenschneider {\it et al.},
	Phys. Rev. Lett. {\bf 86}, 5442 (2001).
\bibitem{Chu.96} L. Chulkov {\it et al.},
	Nucl. Phys. {\bf A603}, 219 (1996).
\bibitem{Cam.01} J. G\' omez del Campo {\it et al.},
	 Phys. Rev. Lett. {\bf 86}, 43 (2001).
\bibitem{PVL.97} W. P\"oschl, D. Vretenar, G.A. Lalazissis,
        and P. Ring, Phys. Rev. Lett. {\bf 79}, 3841 (1997).
\bibitem{LVP.98} G.A. Lalazissis, D. Vretenar, W. P\"oschl,
        and P. Ring, Nucl. Phys. {\bf A632}, 363 (1998).
\bibitem{LVR.97} G.A. Lalazissis, D. Vretenar, W. P\"oschl,
        and P. Ring, Phys. Lett. {\bf B418}, 7 (1998).
\bibitem{Lal.99}  G.A. Lalazissis, D. Vretenar, P. Ring, M. Stoitsov, and
	L. Robledo, Phys. Rev. C {\bf 60}, 014310 (1999).
\bibitem{LVR.98}  G.A. Lalazissis, D. Vretenar, and P. Ring,
	Phys. Rev. C {\bf 57}, 2294 (1998).
\bibitem{VLR.99}  D. Vretenar, G.A. Lalazissis, and P. Ring,
	Phys. Rev. Lett. {\bf 82}, 4595 (1997).
\bibitem{LVR.99}  G.A. Lalazissis, D. Vretenar, and P. Ring,
	Nucl. Phys. {\bf A650}, 133 (1999).
\bibitem{LVR.01b}  G.A. Lalazissis, D. Vretenar, and P. Ring,
	Nucl. Phys. {\bf A679}, 481 (2001).
\bibitem{LVR.01}  G.A. Lalazissis, D. Vretenar, and P. Ring,
	Phys. Rev. C {\bf 63}, 034305 (2001).
\bibitem{Rin.96} P. Ring, Progr. Part. Nucl. Phys. {\bf 37}, 193 (1996).
\bibitem{LKR.97} G.A. Lalazissis, J. K\"onig, and P. Ring,
	Phys. Rev. C {\bf 55}, 540 (1997).
\bibitem{BGG.84}  J. F. Berger, M. Girod and D. Gogny,  
                Nucl. Phys. {\bf A428}, 32 (1984).
\bibitem{AW.95} G. Audi and A. H. Wapstra, 
	Nucl. Phys. {\bf A595}, 409 (1995).
\bibitem{Fuk.91} M. Fukuda {\it et al.},
	Phys. Lett. {\bf B268}, 339 (1991).
\bibitem{Kell.95} J.H. Kelley {\it et al.},
	Phys. Rev. Lett. {\bf 74}, 30 (1995).
\bibitem{Vri.87} H.De Vries, C.W.De Jager and C. De Vries: At. Data and
	  Nucl. Data Tables {\bf 36}, 495 (1987). 
\bibitem{Baz.95} D. Bazin {\it et al.},
	Phys. Rev. Lett. {\bf 74}, 3569 (1995).
\bibitem{Nak.99} T. Nakamura {\it et al.},
	Phys. Rev. Lett. {\bf 83}, 1112 (1999).
\bibitem{Kan.01} Rituparna Kanungo, I. Tanihata, and A. Ozawa,
	Phys. Lett. {\bf B512}, 261 (2001).
\bibitem{MN.95}  P. M\"oller, J.R. Nix, W.D. Myers, and W.J. Swiatecki, At.
	Data Nucl. Data Tables {\bf 59}, 185 (1995).
\bibitem{Ott.89} E.W. Otten, in {\it Treatise on Heavy-Ion Science},
	edited by D.A. Bromley (Plenum, New York, 1989) Vol 8, p. 515.
\bibitem{Keim.00} M. Keim {\it et al.},
	Eur. Phys. J. A {\bf 8}, 31 (2000).
\end{thebibliography}
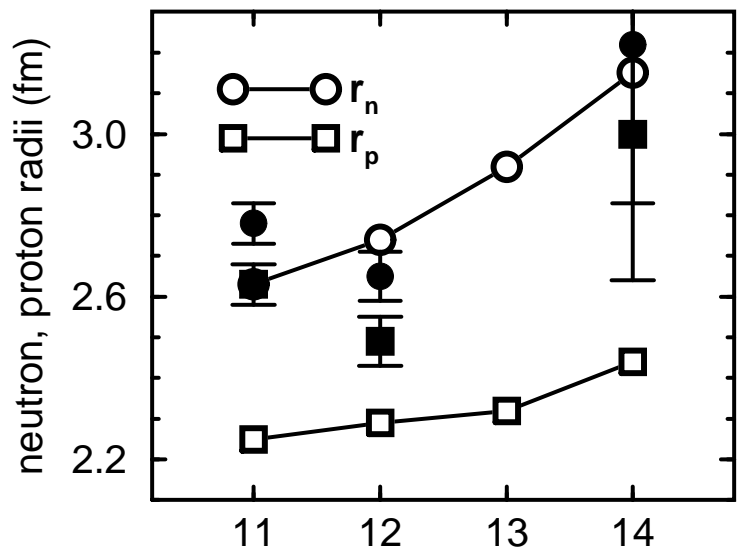
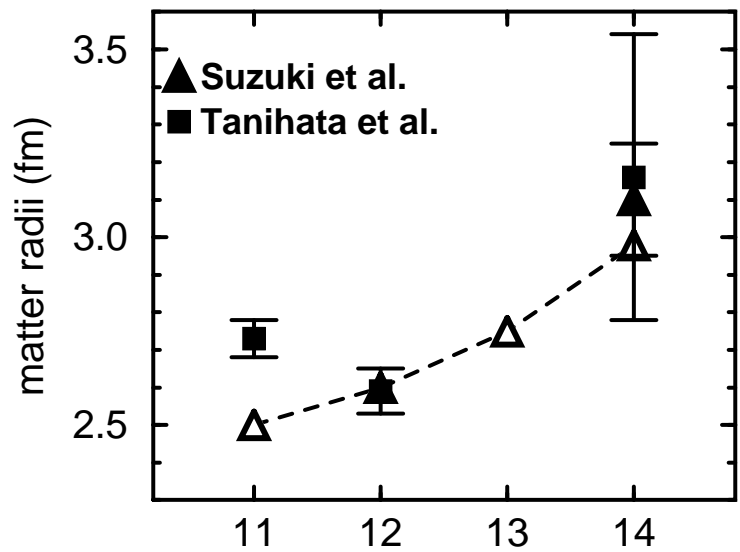
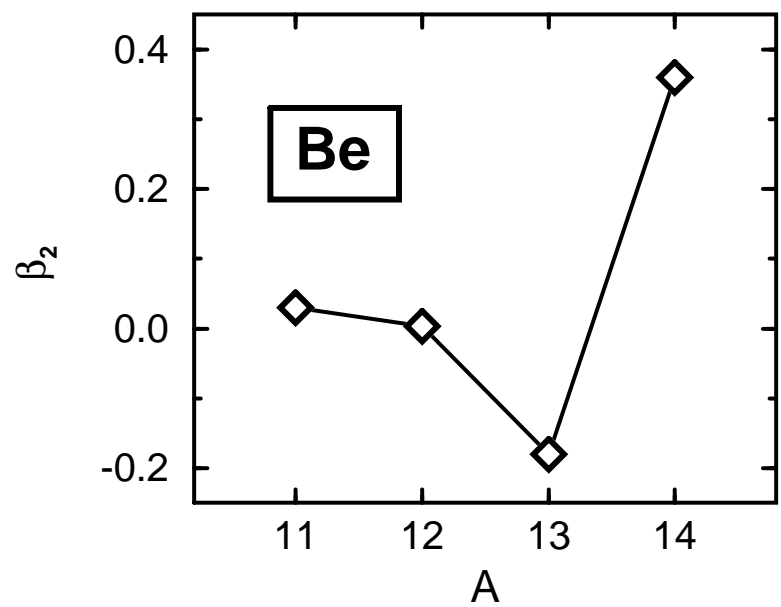
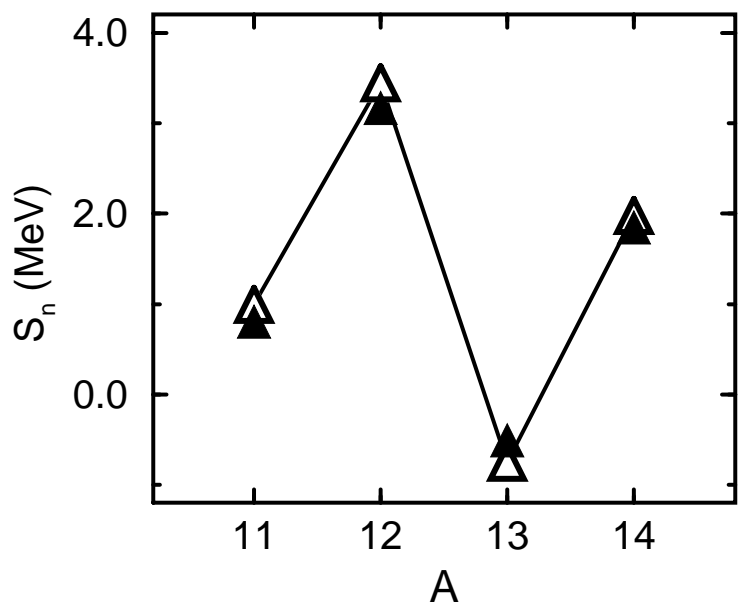
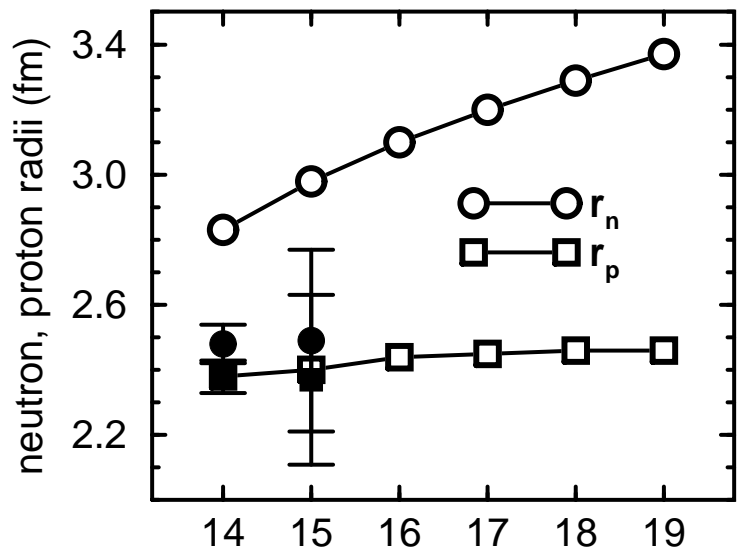
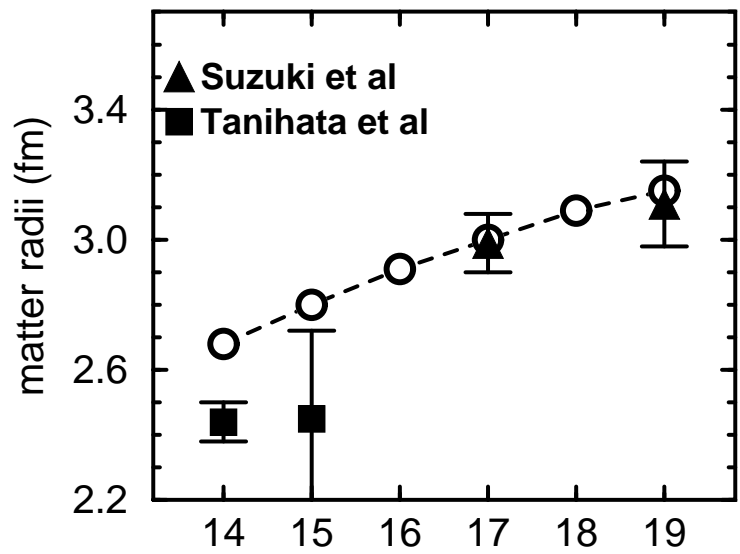
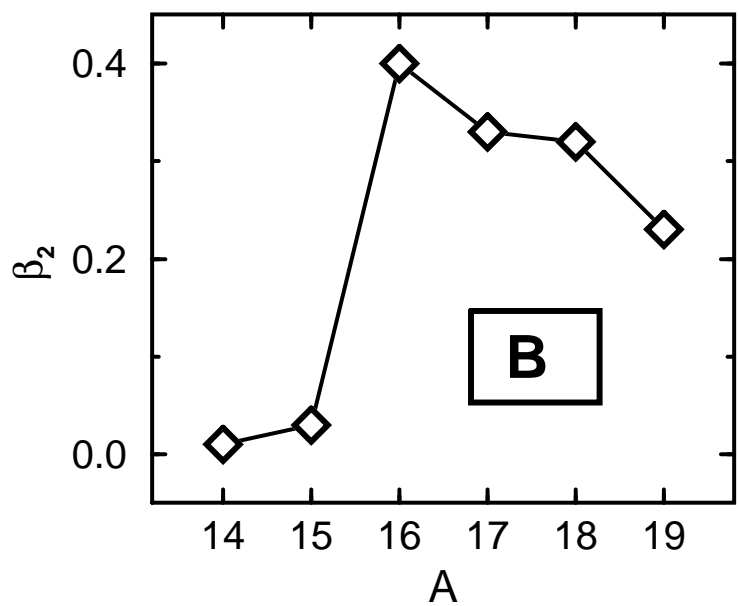
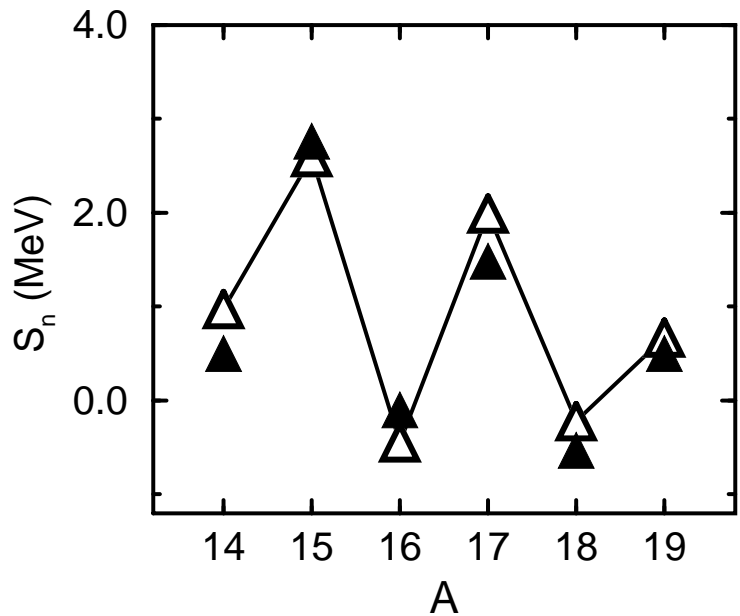
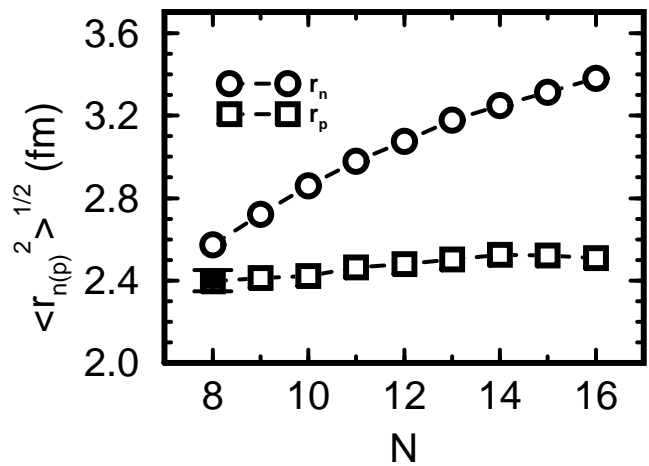
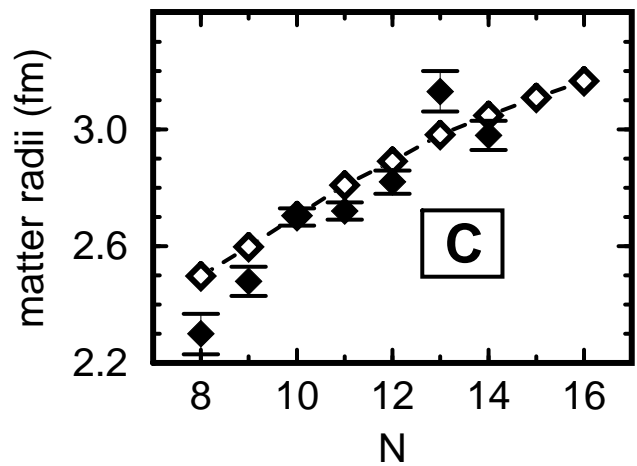
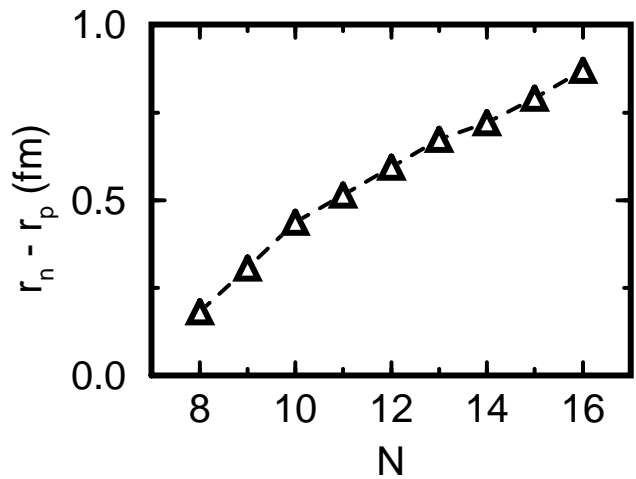
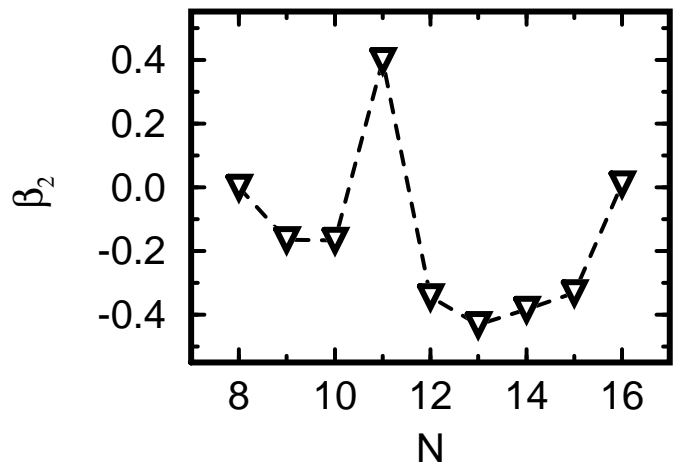
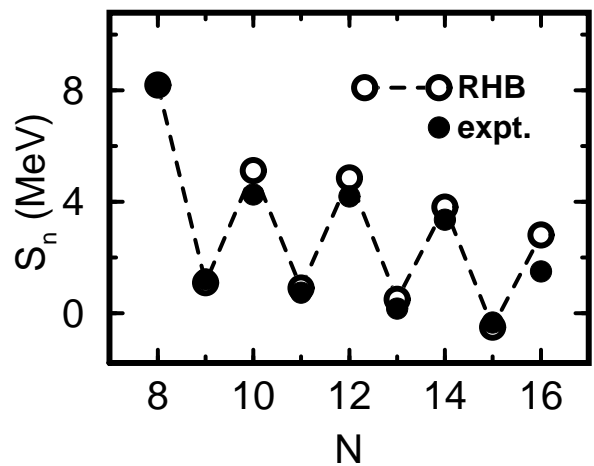
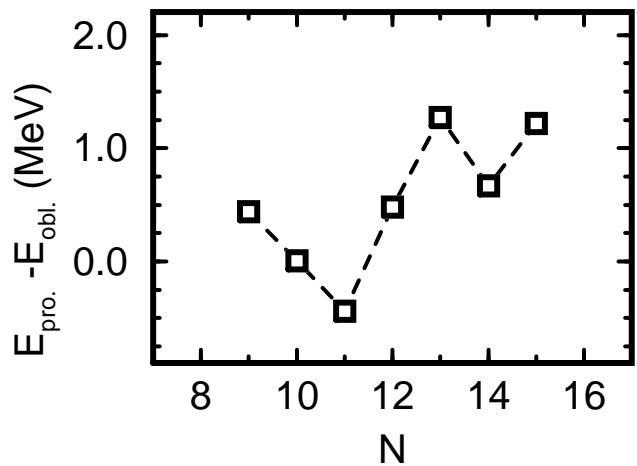
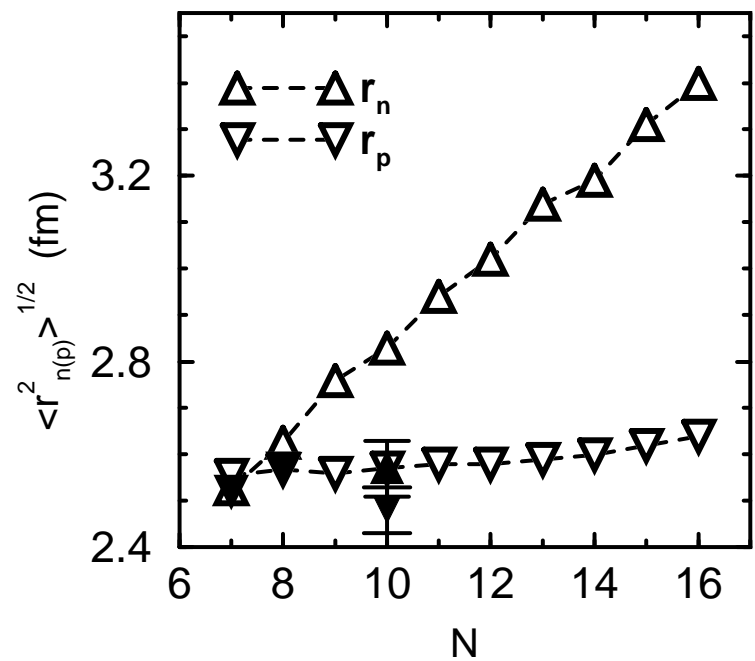
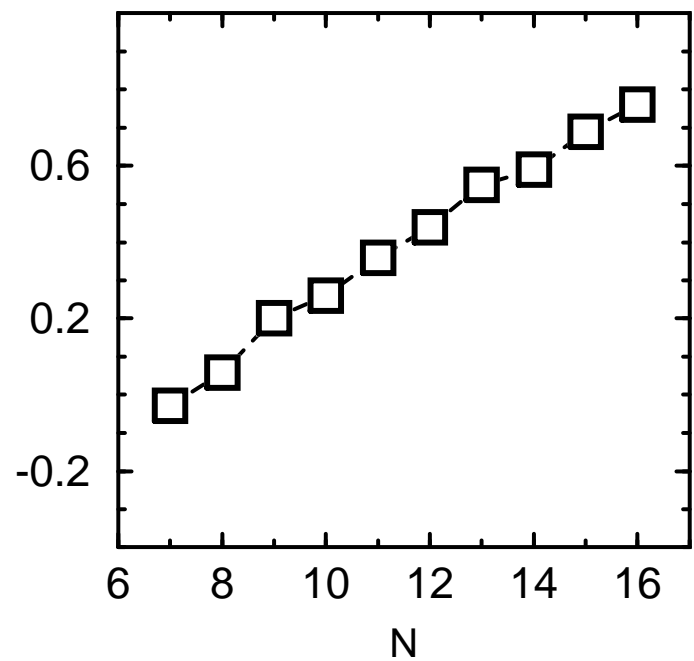
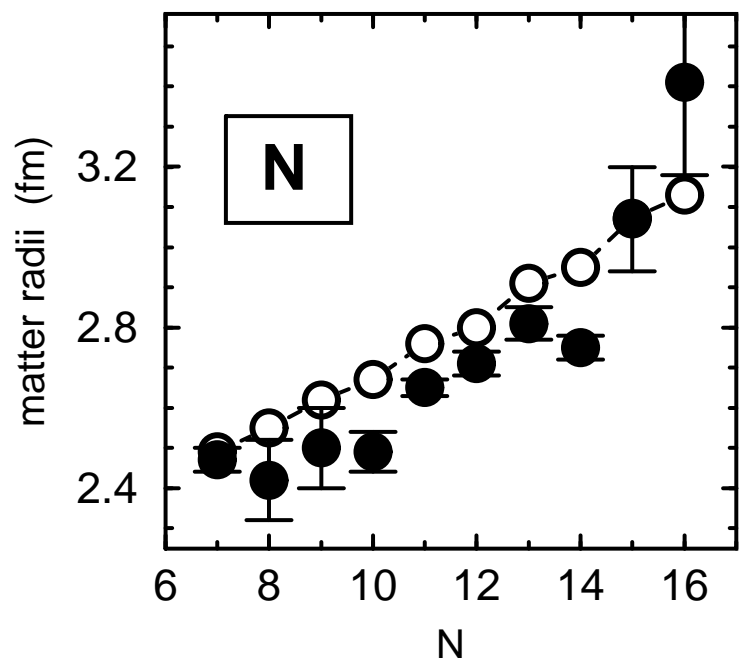
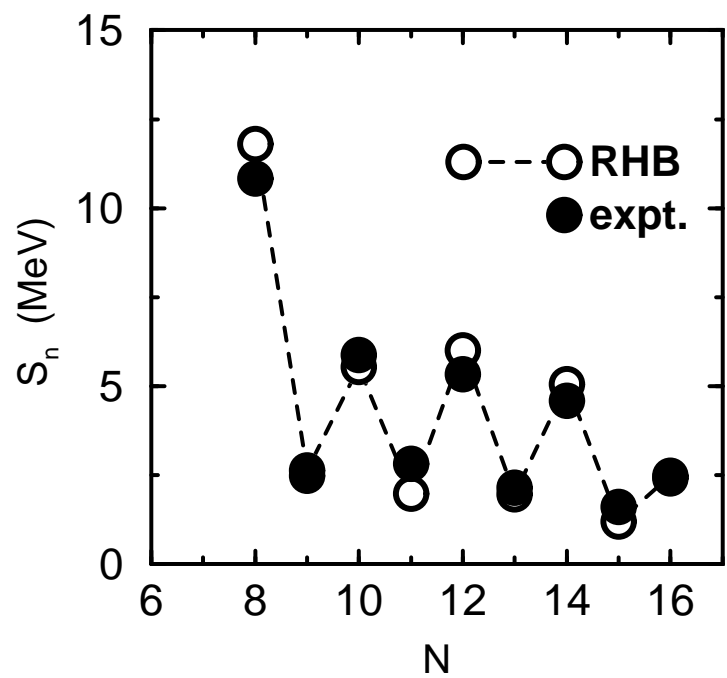
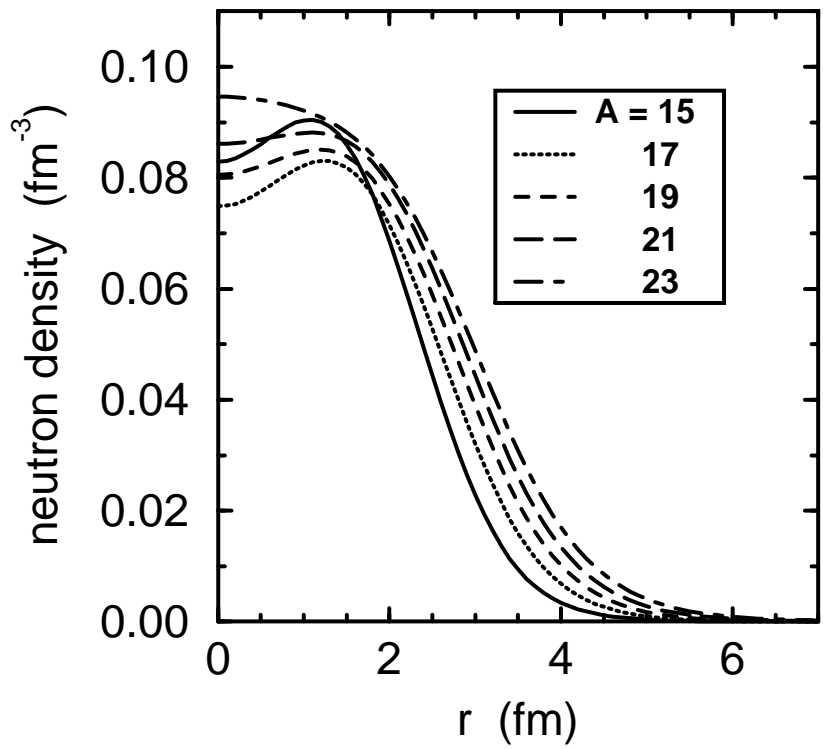
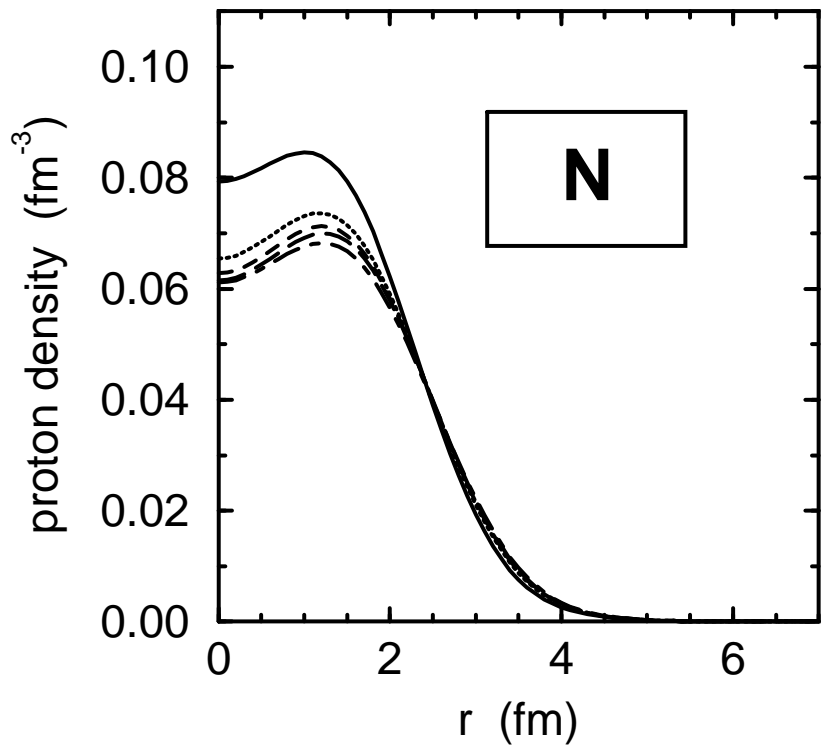
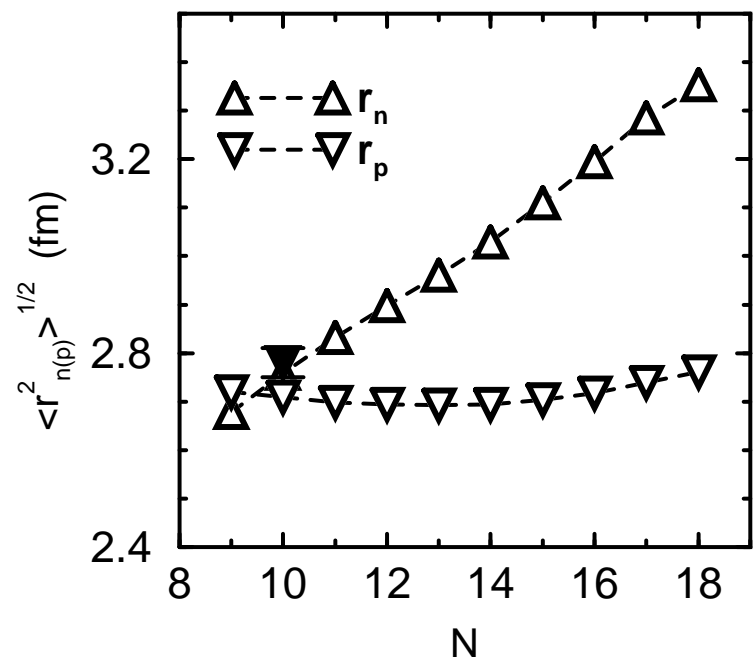
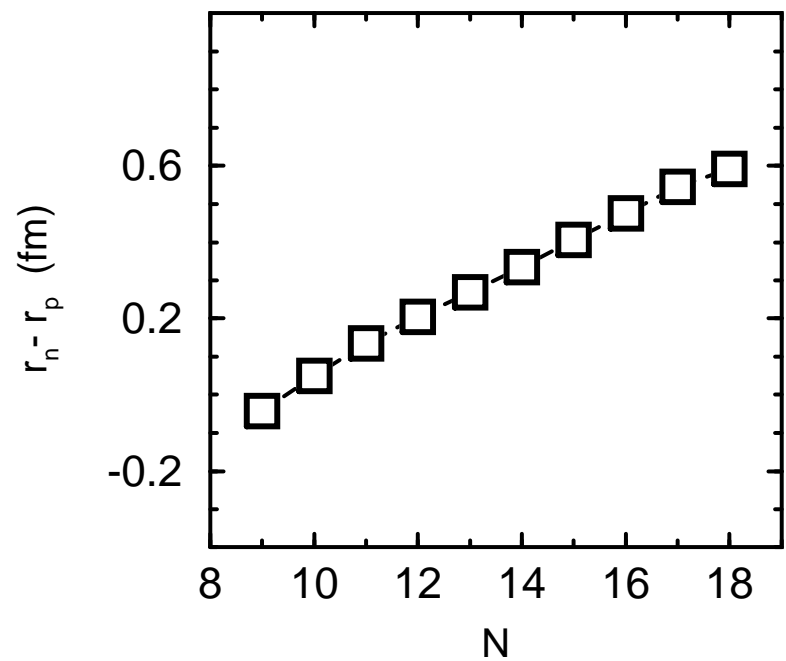
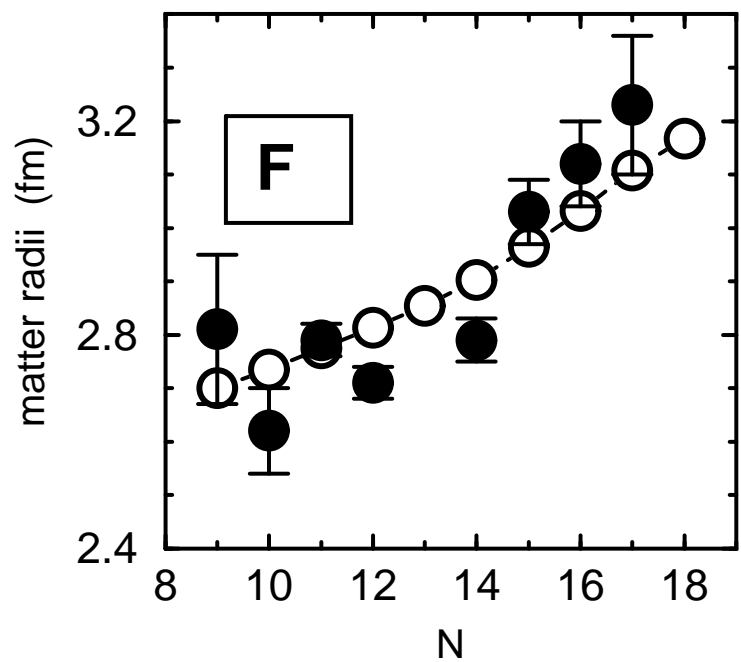
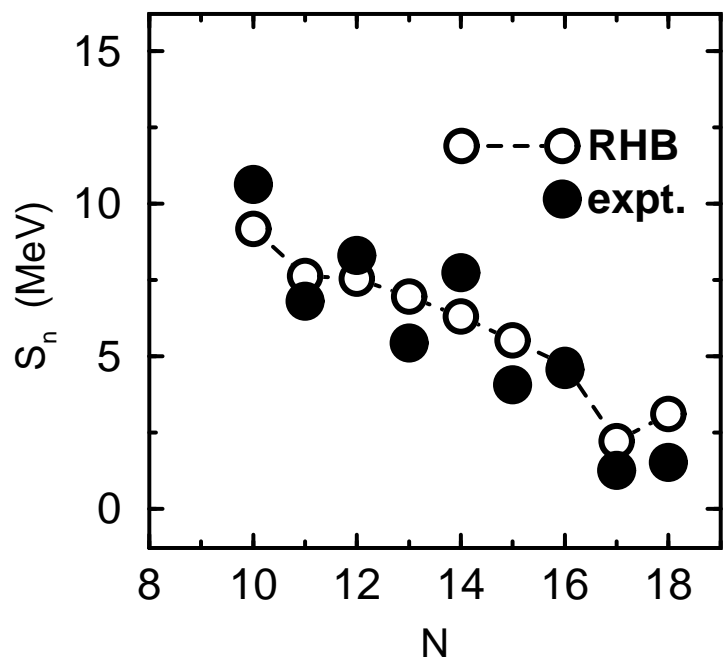
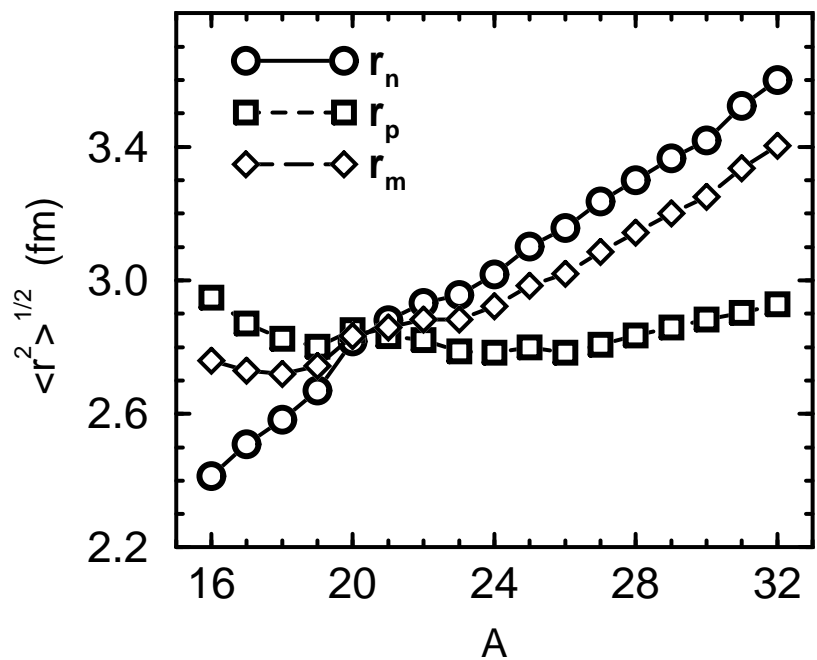
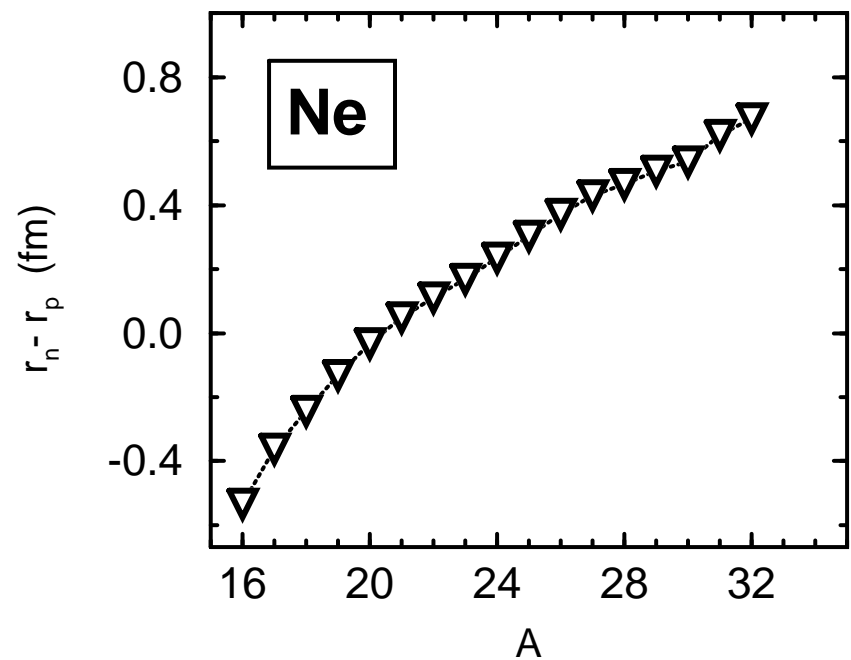
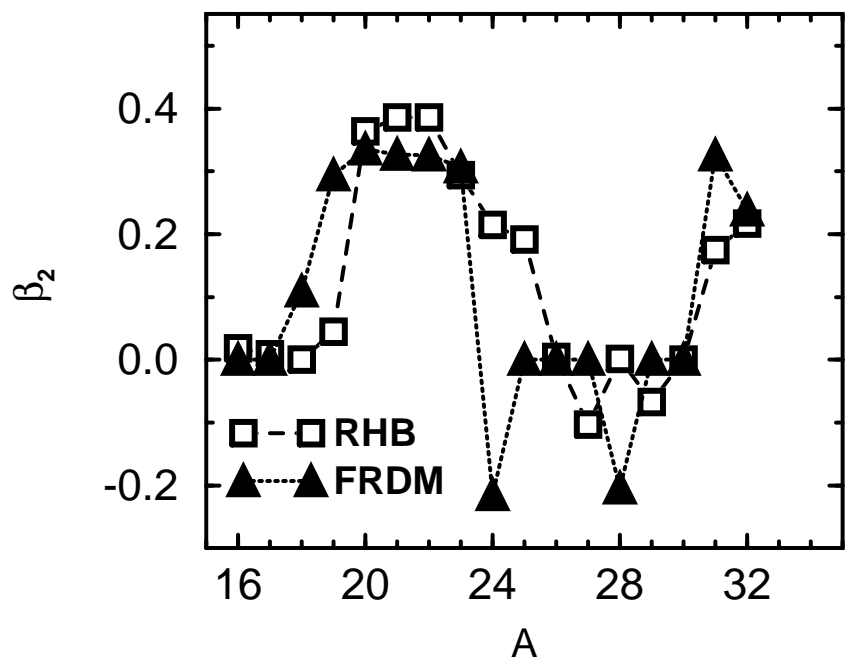
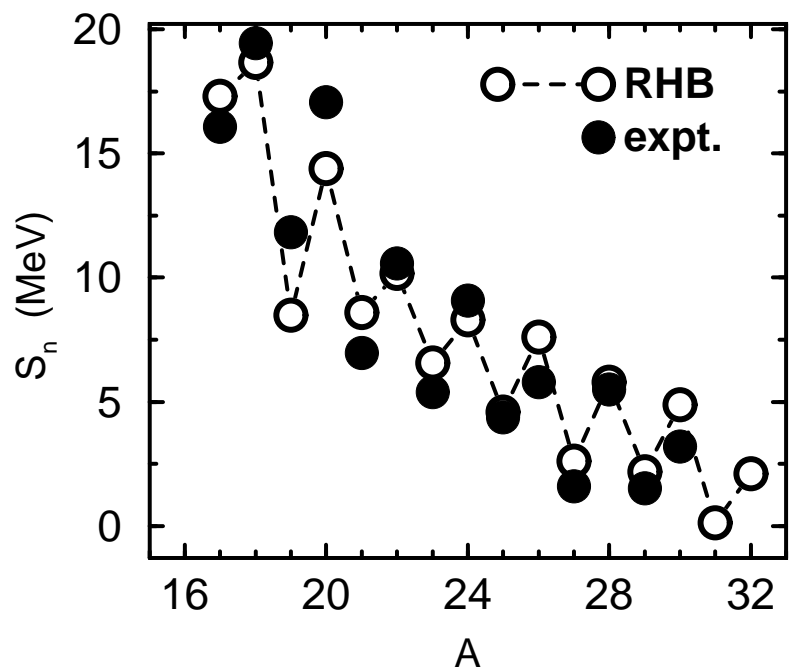
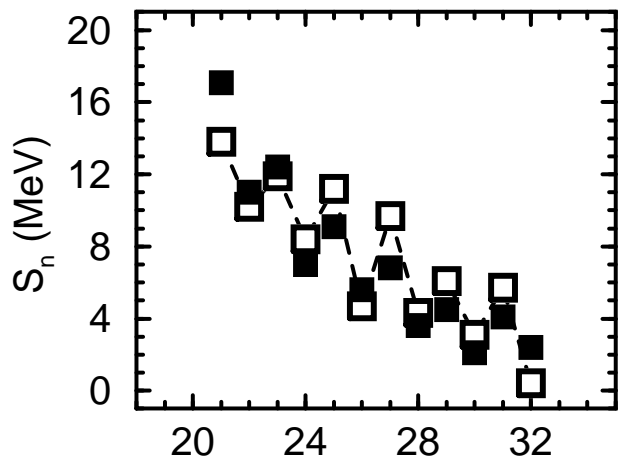
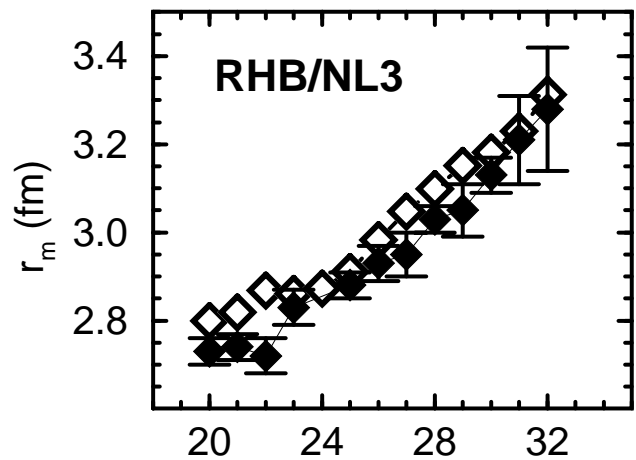
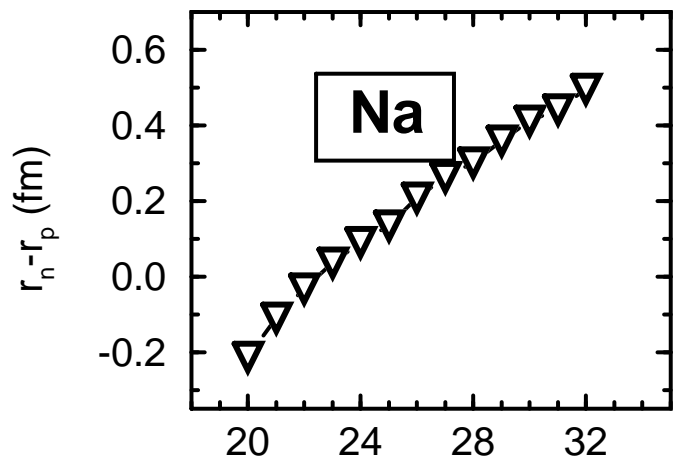
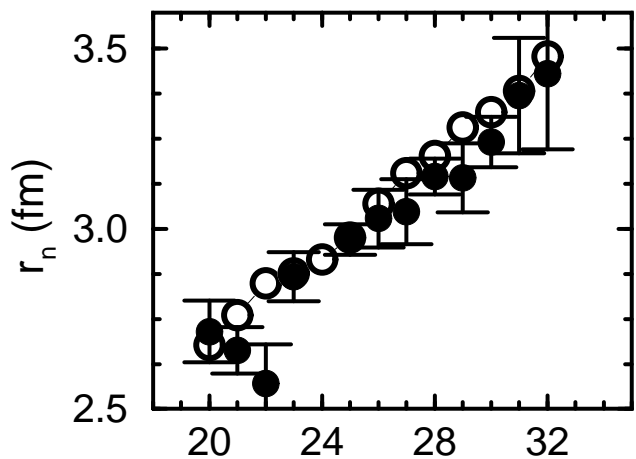
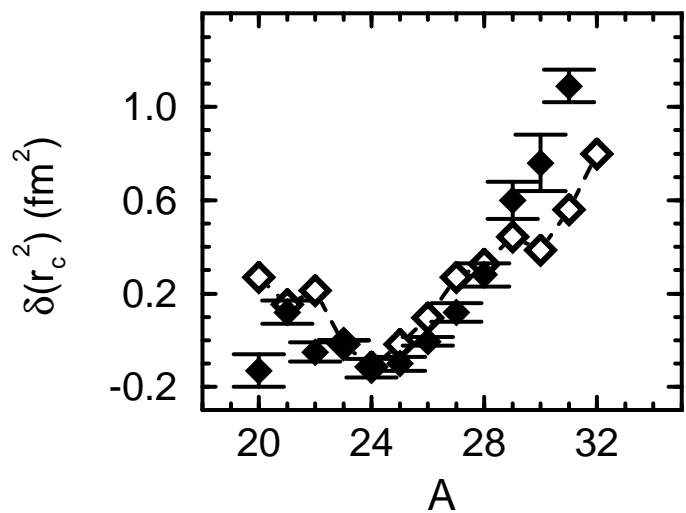
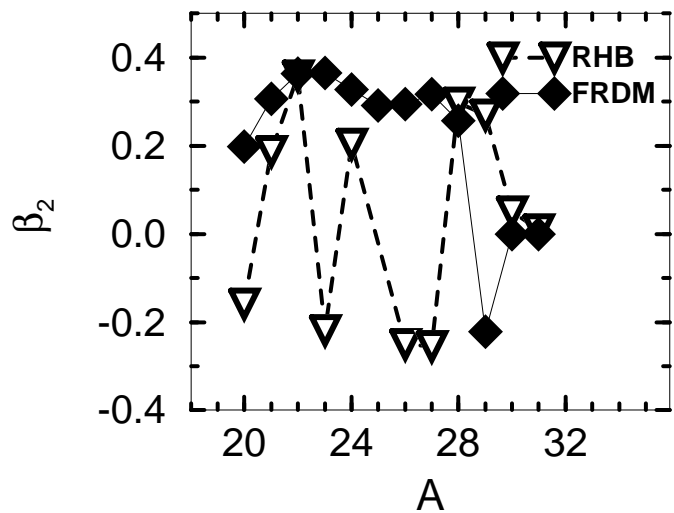
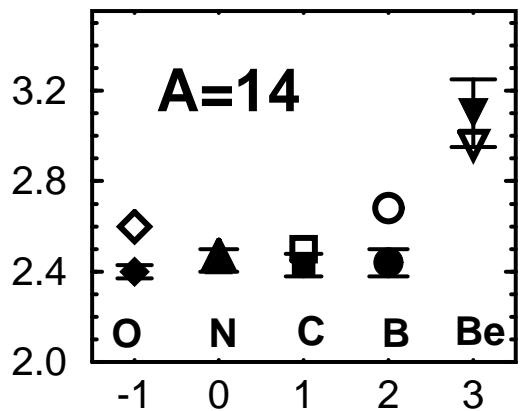
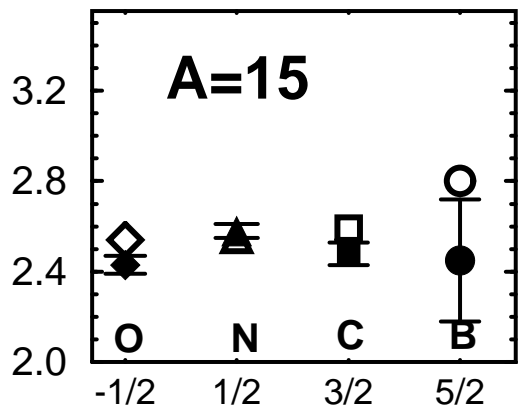
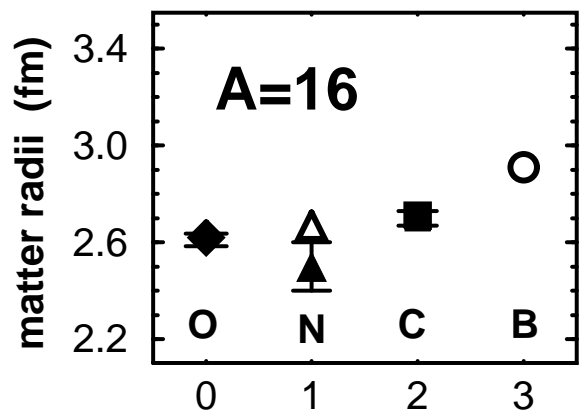
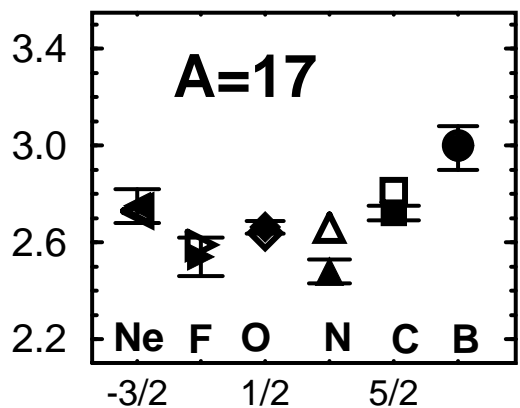
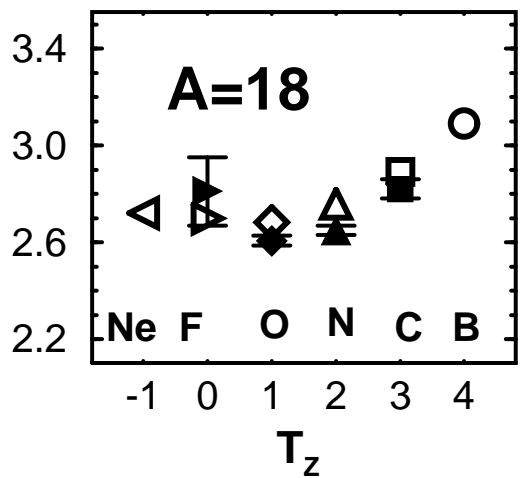
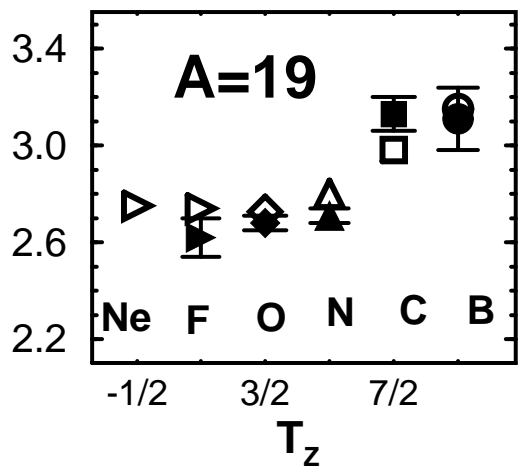
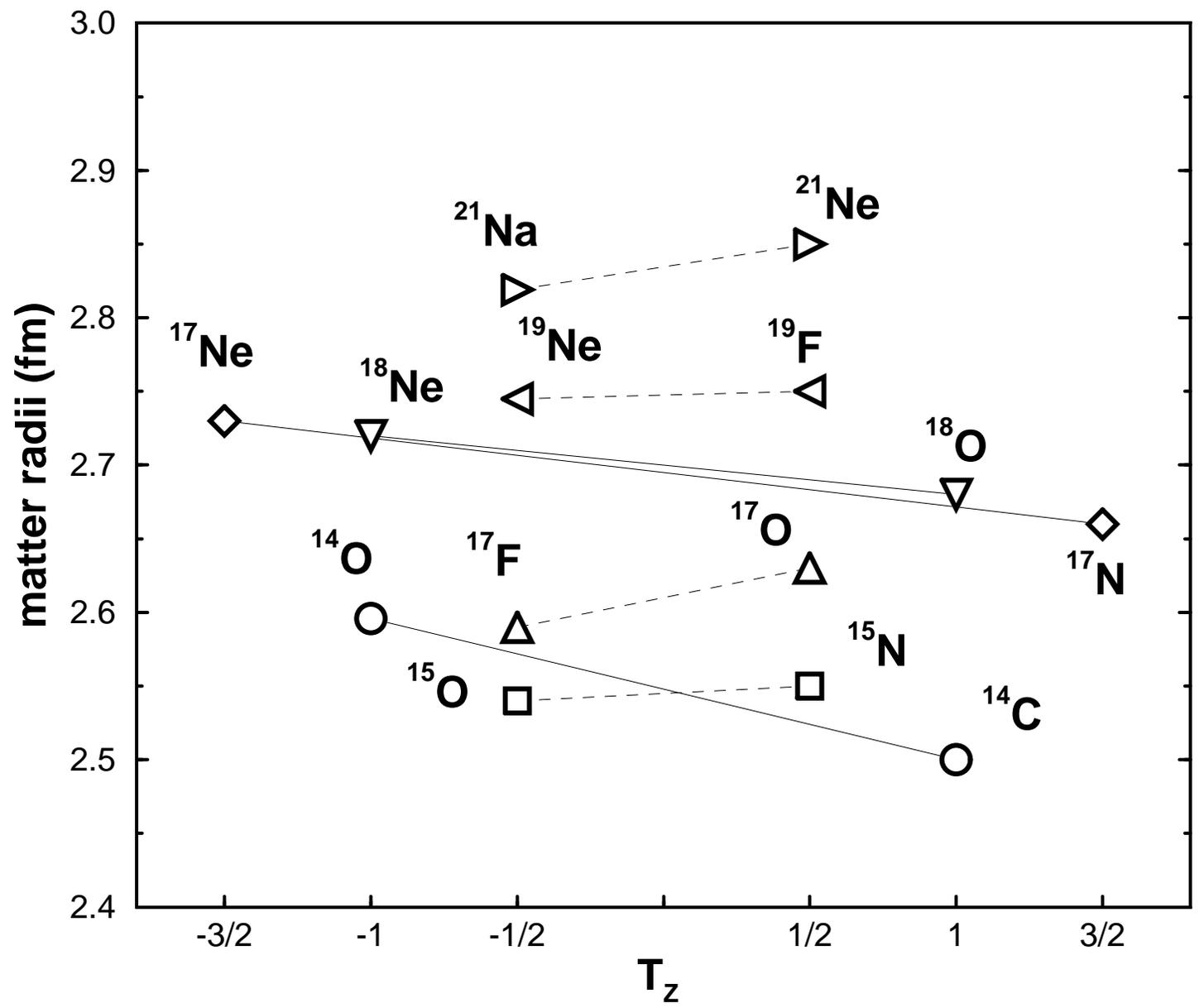
\end{document}